# Automatic Emergency Dust-Free solution on-board International Space Station with Bi-GRU (AED-ISS)

*Po-Han Hou[1]    Wei-Chih Lin[1]    Hong-Chun Hou[2]
Yu-Hao Huang[3]    Jih-Hong Shue[1,4]

1. Department of Space Science and Engineering, National Central University, Taoyuan, Taiwan
2. Department of Atmospheric Sciences, National Central University, Taoyuan, Taiwan
3. Department of Electrical Engineering, National Taipei University of Technology, Taipei, Taiwan
4. Space Environment Laboratory, National Central University, Taoyuan, Taiwan

*Contact: kozak20010716@g.ncu.edu.tw

## Abstract

With a rising attention for the issue of PM2.5 or PM0.3, particulate matters have become not only a potential threat to both the environment and human, but also a harming existence to instruments onboard International Space Station (ISS). Our team is aiming to relate various concentration of particulate matters to magnetic fields, humidity, acceleration, temperature, pressure and $CO_2$ concentration. Our goal is to establish an early warning system (EWS), which is able to forecast the levels of particulate matters and provides ample reaction time for astronauts to protect their instruments in some experiments or increase the accuracy of the measurements; In addition, the constructed model can be further developed into a prototype of a remote-sensing smoke alarm for applications related to fires. In this article, we will implement the Bi-GRU (Bidirectional Gated Recurrent Unit) algorithms that collect data for past 90 minutes and predict the levels of particulates which over 2.5 μm per 0.1 liter for the next 1 minute, which is classified as an early warning.

## 1 Introduction

On December 21, 2021, ICE Cube's AI Box was launched aboard the well-known SpaceX CRS-24 rocket (Figure 1, © AI Space Challenge 2021) on its way to the International Space

Station (ISS). This AI Box was equipped with a set of sensors to monitor various environmental factors and was integrated with the powerful edge computing system, Nvidia Jetson Xavier NX. The sensors inside the AI Box served different purposes. The SparkFun Atmospheric Sensor Breakout (BME280) was responsible for measuring ambient pressure, humidity, and temperature. The Adafruit PMSA003I Quality Breakout (PMSA) was tasked with measuring particulate concentration and number density. Lastly, the Adafruit SCD-30 – NDIR CO2 Temperature and Humidity Sensors (SCD) and the SparkFun 9DoF IMU Breakout (ICM-20948) were utilized for measuring acceleration and magnetic field, respectively.

In addressing environmental and safety concerns, our team is developing an Early Warning System (EWS) capable of forecasting particulate matter concentration onboard the International Space Station (ISS) using various sensors, including acceleration, temperature, humidity, magnetic fields, and CO2 concentration. This idea draws inspiration from previous research, specifically the Dust and Aerosol Measurement Feasibility Test (DAFT) [1][2][3]. We have chosen to incorporate acceleration as it plays a crucial role in particulate attachment to equipment surfaces. Additionally, temperature and humidity have been recognized as factors influencing particulate matter concentration [4]. Considering metal particles generated within the ISS and their potential interaction with the geomagnetic field, we have included magnetic field measurements in our study. Furthermore, based on NI et al.[2017] [5], the presence of metal particles significantly impacts charge distribution on insulating surfaces, leading to a nearly threefold increase in maximum surface charge density compared to the absence of metal particles.

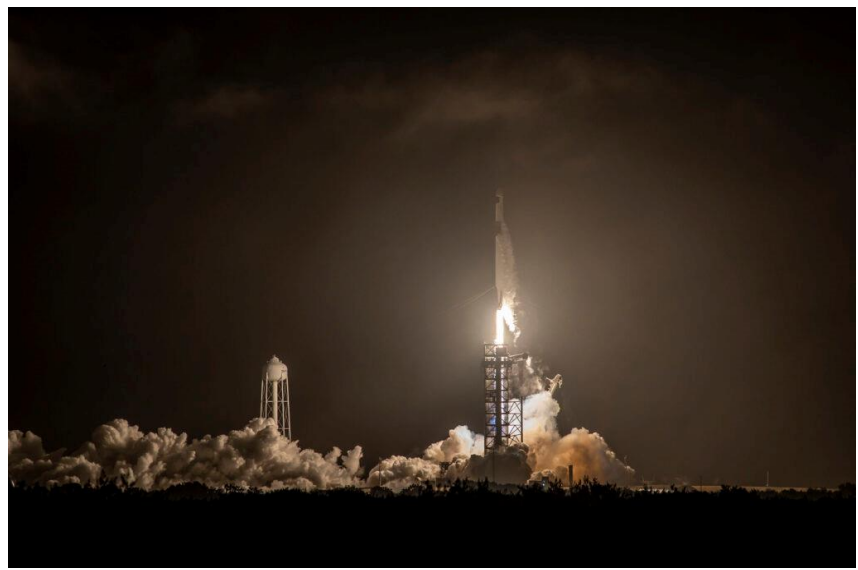

Figure 1. SpaceX CRS-24 lift-off scene

Furthermore, despite maintaining stable pressure and CO2 concentration on the ISS, there might be other factors that have not been considered yet. To enhance the accuracy of our prediction model, we have decided to incorporate these additional factors into our study. Based on scientific hypotheses, we utilize data from the accelerometer, temperature, humidity, magnetic field, pressure, and CO2 concentration as input features in our deep learning algorithm. The target output consists of data for various particulates. Employing the Bi-GRU model, our aim is to develop a predictive model that takes data from these sensors over approximately the past 90 minutes and forecasts the concentration of particulates for the next minute, thereby achieving our research objective.

## 2 Objective

The primary objective of our solution is to develop a surrogate model for the existing Caution & Warning (C&W) System onboard the ISS, which has been operational for several years [6]. To ensure the success of our Early Warning System (EWS) on the ISS, we have identified four crucial requirements. Firstly, we prioritize achieving a high correct rate, as it is fundamental to the efficacy of our EWS concept. Secondly, our model must be capable of issuing warning signals when the prediction of PM2.5 concentration exceeds the threshold specified by the Department of Health in New York State [7]. This threshold indicates that particulate concentrations above 35 µg/m$^3$ can pose a risk to human alveoli. By doing so, our EWS not only safeguards astronauts' alveoli from PM2.5 exposure but also monitors PM0.3 levels to protect sensitive instruments inside the ISS that are susceptible to particulate interference, serving as our third requirement.

Our design incorporates references to the ISO-14644 class 3 clean room standard and takes into account the impact of metal particles on surface charge accumulation [7]. Finally, our model will be retrained on a monthly basis to adapt to potential factors such as the solar maximum, which occurs every 11-year cycle [8] and compresses Earth's magnetic field, affecting the geomagnetic data detected by the AI Box. This regular retraining ensures the continued accuracy and adaptability of the EWS in the dynamic space environment.

## 3 Methodologies

The recognition of air particle interference with precision instruments has been the driving force behind our research on early warning systems. In response, we have chosen to leverage neural networks and artificial intelligence to predict air quality in the upcoming seconds or minutes, granting us vital reaction time to implement preventive measures.
Our team's primary objective is to forecast particle density for the next minute, utilizing data from the past few minutes, including temperature, humidity, acceleration, and magnetic field strength. To achieve this, we intend to employ the Bi-GRU algorithms model, Adam optimizer, and ReLU activation function. Beyond safeguarding precision instruments, this concept holds potential applications in smoke and fire alarms. We are confident that our product could find valuable commercial use in property protection and enhancing social security measures.

## 4 Implementation of Bi-GRU

### 4.1 GRU

The selection of GRU (Gated Recurrent Units) as our primary AI algorithm is attributed to the nature of our model, which primarily focuses on predicting future particle concentrations—a task that involves time series analysis. While considering alternative approaches such as LSTM [9][10][11] and ARIMA, we found that ARIMA's requirement for stable data and its limitation to single-variable prediction make it less suitable for handling the non-linear data prevalent in natural environments. Moreover, LSTM, while effective, is relatively slower and requires more memory compared to GRU [12]. Given the imperative need for fast and efficient prediction to achieve our Early Warning System (EWS) objective, our team opted for the GRU algorithm as it strikes a balance between accuracy and computational efficiency. This choice empowers us to meet the real-time demands of the EWS while effectively predicting future particle concentrations in dynamic environmental settings.

### 4.2 Functional Block Diagram

Of all machine/deep learning projects, there no doubt exists a processing format such as data cleaning, feature engineering, etc. Likewise, the establishment of AED-ISS have the same procedure. First of all, we must determine what we have to do in every phase , and then build a functional block diagram accordingly (Figure 2).

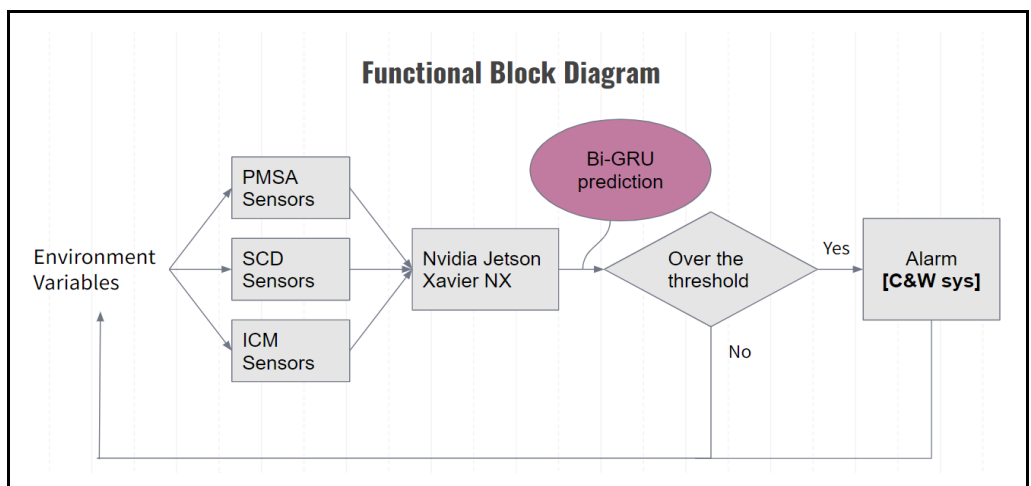

Figure 2, Functional Block Diagram about AED-ISS

The functional block diagram of our system is as follows:
1.  Environmental factors' values are collected by AI Box using sensors such as PMSA (Particulate Matter Sensor), SCD (CO2, Temperature, and Humidity Sensor), and ICM (Accelerometer and Magnetic Field Sensor).
2.  The collected data is then transmitted to Nvidia Jetson Xavier NX, where the predictive model utilizing the Bi-GRU algorithm processes the information.
3.  The model determines whether the concentration of PM2.5 or PM0.3 exceeds the predefined threshold.
4.  If the concentration surpasses the threshold, an alarm signal is promptly sent to the onboard Caution and Warning System on the ISS to alert astronauts of potential risks.
5.  In case the concentration is below the threshold, the system continues the loop, continuously monitoring and predicting air quality.

By following this functional block diagram, our Early Warning System effectively detects and notifies the ISS crew of hazardous particulate levels, enabling them to take timely measures to protect both precision instruments and their health in response to environmental fluctuations.

### 4.3 Data Extraction

Our team has designed a particulate Early Warning System (EWS) that utilizes data from the AI Box, including particle concentration, pressure, CO2, humidity, temperature, acceleration, and magnetic field. To train and test the model, we aim to collect 24 hours of data from the AI Box. The data collection will begin before 00:00:00Z, and by 24:00:00Z, the model will undergo training and testing. After training, we will use the time interval between 22:30:00Z and 00:00:00Z to predict the particle concentration for the next one minute, specifically at 00:01:00Z. We have chosen a time interval of ninety minutes because it aligns with the orbital period of the ISS, which is approximately ninety minutes.

To ensure the model's effectiveness and adaptability, we will update it once a month. This update will help maintain consistency between the training and prediction data, ensuring that the model accurately predicts future particulate concentrations, supporting the successful implementation of the Early Warning System on the ISS.

### 4.4 Feature Engineering

After a thorough observation of all the data, we took several essential steps to prepare it for analysis. Firstly, we identified and removed outliers, as well as eliminated any NaN values, ensuring that the remaining data points maintained continuity in the time series. Next, to optimize the dataset, we conducted coefficient correlation tests to eliminate redundant data.

However, a significant challenge arose due to the substantial proportion of 0 values in the datasets from the PM sensors. To address this issue, we employed the under-sampling method to tackle the problem of imbalanced data, allowing us to mitigate the impact of the dominant 0 values.

Moreover, to streamline the data for further analysis, we processed multiple data points recorded within a single second. By averaging them together, we transformed the data into one-second intervals. This approach ensures consistency and enhances the efficiency of our subsequent analysis and modeling efforts.

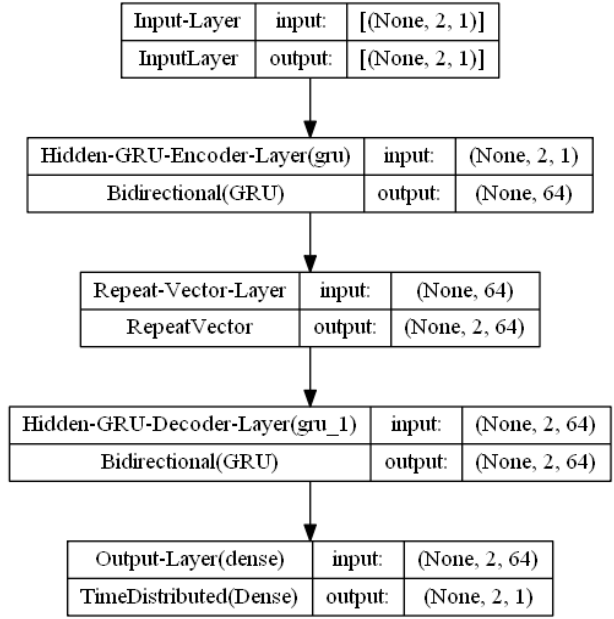

Figure 3. Composition of Bi-GRU layers.

Upon completion of data preprocessing, we proceed to normalize the data to a scale of 0 to 1, which allows for feature transformation onto a standardized scale. Subsequently, we utilize 24 hours of data to generate our model, partitioning it into 65% for training, 10% for validation, and 25% for testing purposes.

### 4.5 Construction of Bi-GRU[13][14][15]

To address the many-to-many prediction nature of AED-ISS, as depicted in Figure 3, we have adopted an encoder-decoder configuration frequently utilized in Natural Language Processing. Both the encoder and decoder consist of hidden GRU layers, facilitating information transfer between them through a repeat vector layer positioned between the encoder and decoder. Furthermore, we incorporated a Bi-Directional wrapper into our model, enabling bi-directional training, which has demonstrated improved outcomes in certain cases. To ensure the proper input shape for decoder layers, we added a repeat vector. Lastly, we employed a time-distributed wrapper to serve as the outputs of AED-ISS, enabling individual predictions for each timestep.

## 5 Results and Analyses

### 5.1 Loss

For our AED-ISS model, we leverage the advantages of ReLU activation function and Adam optimizer. ReLU is chosen for its effectiveness in mitigating the gradient explosion problem, while Adam offers fast deployment and self-adaptive capabilities. The loss graph depicting the model's performance is shown below (Figure 4).

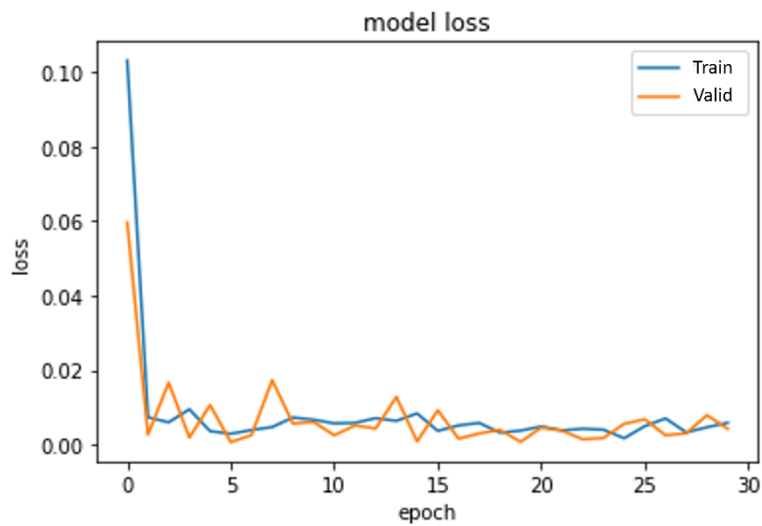

Figure 4. Model loss

Upon generating the model, we plotted the loss graph to evaluate its training progress. Notably, the train loss graph exhibited a significant descending slope, indicating a successful fitting of the data during the training phase. After several epochs, both the train loss and validation loss approached saturation, with a minimal gap between them. This observation suggests that the model's performance achieved a stable and satisfactory level of convergence during training.

### 5.2 Visualization of predicting result

We reserved 25% of the data as test datasets to evaluate our model's accuracy. The results indicate a promising potential for deploying AED-ISS on the ISS to effectively function as an Early Warning System (EWS). The prediction results for AED-ISS in predicting PM0.3 are shown below.

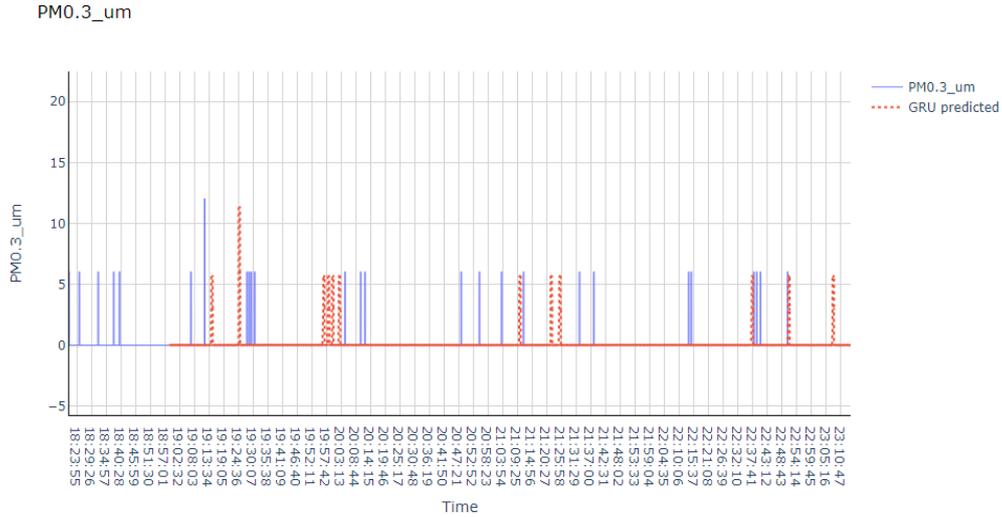

Figure 5. Predicting result of PM0.3

(Blue line represents the true value, and red dot line represents the prediction.)

In real-world data, conditions are rarely perfect for data recording. When predicting PM0.3, we observe that the prediction closely resembles the real data, although the accuracy may fall slightly short of perfection. Despite this, we find the results to be promising and encouraging. The model's ability to closely resemble the actual data suggests that it holds potential for effective application in practical scenarios. While the accuracy may not be absolute, we consider the performance promising enough to be a valuable tool in real-world situations.

### 5.3 RMSE test[16]

The "root mean square error" (RMSE) serves as a measure of the standard deviation of prediction errors or residuals. It quantifies the spread of these residuals, and a lower RMSE indicates a better model performance. Table 1 displays the results of the RMSE test comparing ordinary GRU and Bi-GRU models. The findings reveal that the Bi-GRU model outperforms the ordinary GRU model, demonstrating the superior effectiveness of Bi-GRU in our analysis.

Table 1.

| Model | RMSE |
| --- | --- |
| Ordinary GRU | 0.879 |
| Bidirectional GRU | 0.532 |

# 6 Conclusion and Future works

Our research topic, "Automatic Emergency Dust-Free solution on-board International Space Station (AED-ISS)," draws inspiration from NASA's previous research, DAFT. Given the increasing public concern about particulates, particulate matters have become a potential threat. Our primary objective is to establish an Early Warning System (EWS) capable of forecasting particulate matter levels, providing astronauts with sufficient reaction time to protect their instruments during experiments or improve measurement accuracy. To achieve this, we utilize magnetic field, acceleration, and CO2 concentration data to predict PM2.5 and PM0.3 levels. In contrast to the ordinary GRU module, our team has integrated concepts from NLP to develop a Bidirectional GRU model, which significantly enhances the accuracy of our predictions.

The results demonstrate that the original GRU module consistently predicts zero, whereas the modified bidirectional GRU predictions closely resemble the actual values. Furthermore, we validated these results using the RMSE metric, confirming that the Bi-GRU model outperforms the original one, affirming its potential effectiveness on the ISS.

Additionally, we envision that countries with forests or challenging environments could potentially apply this model with AI Box to function as a wildfire detector. This application would not only prevent people from suffering environmental crises but also protect wildlife habitats from devastating losses. With the encouraging results and dedicated efforts, we firmly believe in developing a new generation of Early Warning Systems that can make a positive impact on various environmental challenges.

# Acknowledgements

Major thanks to the AI Space Challenge committee for providing ICE Cube's AI Box data, and awarding us the Best Technical Award in the challenge. Our team want to thank the CEO of Gran Systems, Ke-Kuang Han, and Professor of Taipei-Tech, Yang-Lang Chang for meaningful discussions and support. We gratefully appreciate Space Environment Laboratory, National Central University and Taipei-Tech for providing edge-computing system to simulate Nvidia Jetson Xavier NX onboard ISS.